# Proton-transfer Ferroelectricity / Multiferroicity in Rutile Oxyhydroxides

Menghao Wu[1*], Tianci Duan[1], Chengliang Lu[1], Huahua Fu[1], Shuai Dong[2], Junming Liu[3]

[1]School of Physics and Wuhan National High Magnetic Field Center, Huazhong University of Science and Technology, Wuhan 430074, China

[2]Department of Physics, Southeast University, Nanjing 211189, China

[3]National Laboratory of Solid State Microstructures, Nanjing University, Nanjing 210093, China

**Abstract**

Oxyhydroxide minerals like FeOOH have been a research focus in geology for studying the Earth's interior, and also in chemistry for studying oxygen electrocatalysis activity. In this paper we provide first-principles evidence of a new class of ferroelectrics/multiferroics among them: β-GaOOH, InOOH, β-CrOOH, ε-FeOOH, which are earth-abundant minerals and have been experimentally verified to possess distorted rutile structures, are ferroelectric with considerable polarizations(up to 24 μC/cm$^2$) and piezoelectric coefficients. Their atomic-thick layer may possess vertical polarization robust against depolarizing field due to the formation of O-H…O bonds that can hardly be symmetrized. Moreover, β-CrOOH (guyanaite) is revealed to be a combination of high-$T_c$ in-plane type-I multiferroics and vertical type-II multiferroics, which is strain-tunable and may render a desirable coupling between magnetism and ferroelectricity. Supported by experimental evidence on reversible conversion between metal oxyhydroxides and dioxides and their nice lattice match that renders convenient epitaxial growth, heterostructure composed of oxyhydroxides and prevalent metal dioxides (e.g. TiO$_2$, SnO$_2$ and CrO$_2$) may be constructed for various applications like ferroelectric field-effect transistors and multiferroic tunneling junctions.

Oxyhydroxide minerals such as AlOOH[1] and FeOOH[2] in the Earth's mantle have been a research focus of geologists for studying the water storage capacity of the Earth's interior, and of chemists for studying oxygen electrocatalysis activity.[3-7] They have also attracted considerable interest from physicists partially for studying the symmetrization of hydrogen bonds under high compression, which may have significant effect on the crystal structures and physical properties. This phenomena have been predicted or even demonstrated in various type of materials like formic acid[8] and ice-X[9], and recently the pressure-induced hydrogen bond symmetrization has been characterized by evidence from a combination of various spectroscopy in oxyhydroxides like AlOOH[10], FeOOH[11] and also CrOOH[12]. It is worth mentioning that many oxyhydroxide polymorphs have been verified. For example, there are three forms of CrOOH, denoted as α-CrOOH (grimaldiite), ß-CrOOH (guyanaite), and Γ-CrOOH (bracewellite), in which guyanaite is a commonphase mineral with a distorted rutile structure. Some other oxyhydroxides like β-GaOOH and InOOH share similar structures[13, 14], which are usual byproducts in semiconductor industry.

In this paper we focus on an important property that has been scarcely noticed in oxyhydroxide minerals. Although the antiferromagnetism in CrOOH and FeOOH have been investigated in some reports[11, 12], their possible ferroelectricity due to the breaking of inversion symmetry by hydrogen bonds has not yet been explored. It is known that ferroelectric (FE) materials[15-18], which possess spontaneous electric polarizations switchable under external electric field, have a wide range of potential applications in electronics, micromechatronics and electro-optics. Ferromagnetic (FM) materials with switchable magnetization, FE materials with switchable electrical polarizations and ferroelastic materials with switchable strain[19, 20] can find applications in non-volatile memory. In commercial random access memories (RAMs), data writing in FM RAMs is energy-consuming, while reading operation in FE RAMs is destructive. To resolve both issues, multiferroic materials with both FM and FE properties are sought due to the combination of both efficient writing and less energy-cost reading[21]. Due to the challenge in incorporation of both orderings in the same compound, their existences in nature are rare and their Curie temperature are usually far below room temperature[22-25]. Almost all the realized multiferroic materials reported to date are either antiferromagnetic or ferrimagnetic, except $EuTiO_3$ which becomes FM under large strain[26, 27]. They can be classified as either type-I or type-II multiferroics, where FE is induced by magnetism in type-II multiferroics with strong magnetoelectric coupling favorable for efficient data reading and writing.[23] Recently a series of 2D type-I multiferroics with almost independent FE and magnetism have been theoretically predicted[20, 28-32], but are still yet to be synthesized in experiments.

Herein we focus on distorted rutile-type oxyhydroxides[1, 2, 10-14, 33-35], including β-GaOOH, InOOH, β-CrOOH, ε-FeOOH, where the asymmetric O-H…..O configuration at ambient pressure can result in proton-transfer FE. Compared with conventional FE materials, it has already been concluded that proton-transfer FE have many advantages[24, 32, 36]: The steric hindrance or high energy barriers during switching can be avoided, polarity can be formed spontaneously for the directional preference of hydrogen bonding, and strong hydrogen-bonds may also result in high-temperature FE. Now we further demonstrate another advantage by first-principles calculations: their atomic-thick thin-film can exhibit vertical FE robust against depolarizing field, while the potential applications of traditional FE thin films are hindered by their polarization that will disappear below critical film thickness[37, 38]. Moreover, the coexistence of room-temperature magnetism and proton-transfer FE, type-I and type-II multiferroicity, are predicted in β-CrOOH (guyanaite). Due to lattice match, heterostructures composed of different oxyhydroxides and metal dioxides can be constructed for various applications like FE field-effect transistors, multiferroic tunneling junctions (MTJs)[39], or simply for enhanced FE upon epitaxial strain.

**Results and Discussions**

The geometric structures of β-GaOOH, InOOH, β-CrOOH, and ε-FeOOH are displayed in Fig. 1(a), where the hydrogen-bonding geometry of those rutile-type structures leads to reduction of symmetry to space group $Pmn2_1$, which have been already verified in experiments[11, 12, 34]. Here proton-transfer FE may stem from the asymmetric O-H…..O configuration, as shown in Fig. 1(b), where the polarization can be switched upon the hopping of protons along hydrogen bonds. There are two nearly independent O-H…..O hydrogen bonds per unitcell aligned in two directions that are almost perpendicular, so two different types of FE may emerge upon various combinations: switching from I to II along the –Z axis, or from III to IV along the –Y axis, with different direction and magnitude of polarizations. To our calculations, for GaOOH, InOOH and FeOOH, the state I/II are slightly lower in energy compared with state III/IV, which is vice versa for CrOOH. So I/II and III/IV may sometimes be regarded as degenerate considering their negligible energy difference (~meV/f.u.). As summarized in Table 1, GaOOH possesses the largest polarization (~24 µC/cm$^2$) in –y direction. If their metal ions like Fe or Cr possess magnetism, the systems can be multiferroic.

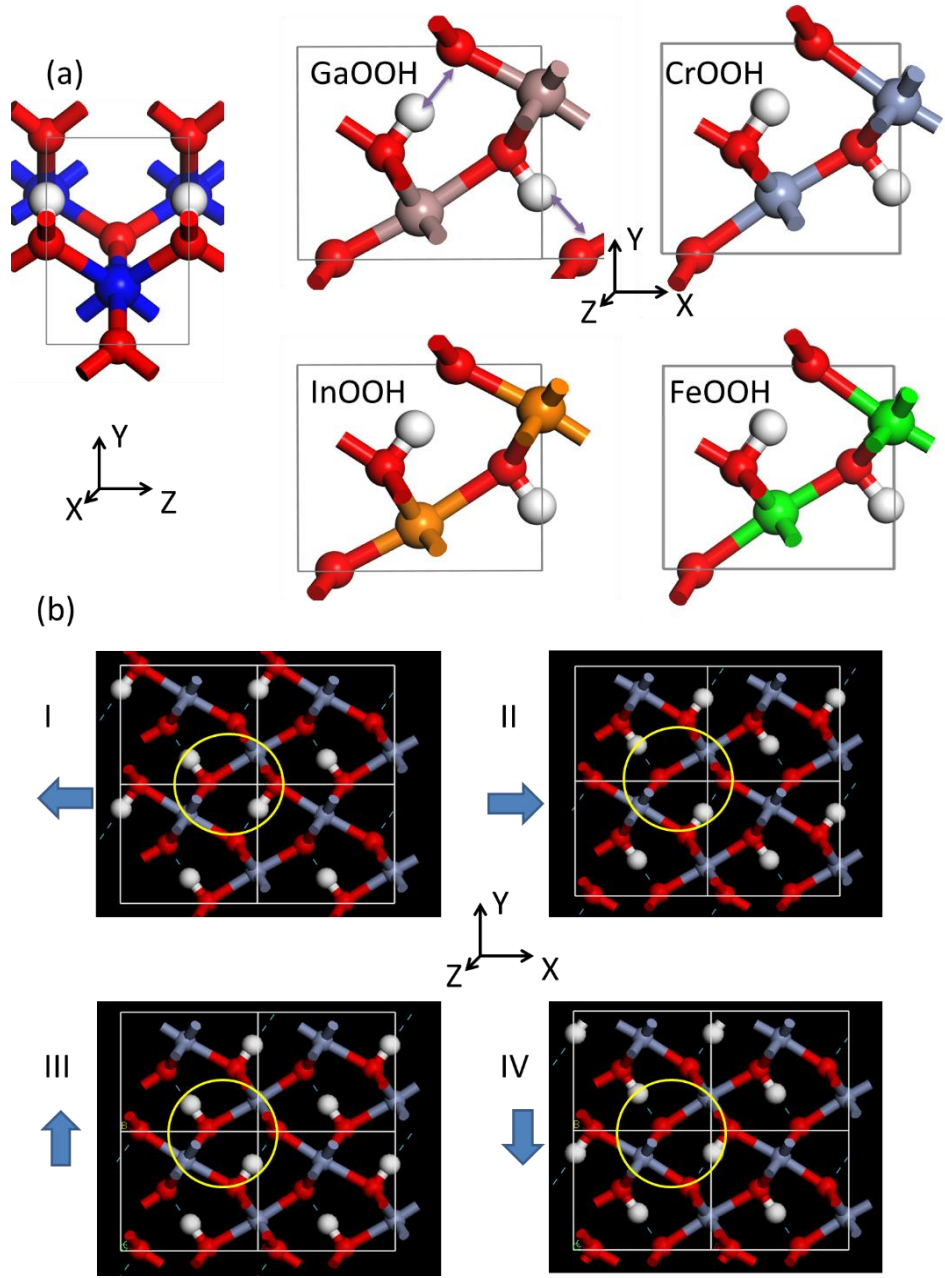

Figure 1 (a)Geometric structures of β-GaOOH, InOOH, β-CrOOH, and ε-FeOOH. (b)Two distinct types of FE switching from I to II and from III to IV, where blue arrows on the left denote the direction of polarizations of I-IV configurations, and purple arrows denote the directions of proton transfer. White, red, pink, blue, orange, green spheres denoted H, O, Ga, Cr, In, Fe atoms respectively.

Table 1 Lattice constants, polarizations and energy difference between state I/II and state III/IV.

|  | \|a\|(Å) | \|b\|(Å) | \|c\|(Å) | $P_x$(μC/cm²) | $P_y$(μC/cm²) | $\Delta E$(meV/f.u.) |
|---|---|---|---|---|---|---|
| **β-GaOOH** | 4.92 | 4.34 | 3.00 | 17.5 | 23.8 | 4.6 |
| **InOOH** | 5.33 | 4.59 | 3.32 | 7.9 | 17.4 | 3.0 |
| **β-CrOOH** | 4.88 | 4.32 | 2.98 | 10.9 | 20.5 | -2.6 |
| **ε-FeOOH** | 4.93 | 4.40 | 3.00 | 23.1 | 20.3 | 12.5 |

Table 2 Lattice constants of rutile metal dioxides.

|  | $TiO_2$ | $SnO_2$ | $CrO_2$ | $RhO_2$ | $RuO_2$ |
|---|---|---|---|---|---|
| \|a\|(Å) | 4.65 | 4.83 | 4.49 | 4.56 | 4.54 |
| \|c\|(Å) | 2.97 | 3.24 | 2.98 | 3.13 | 3.14 |

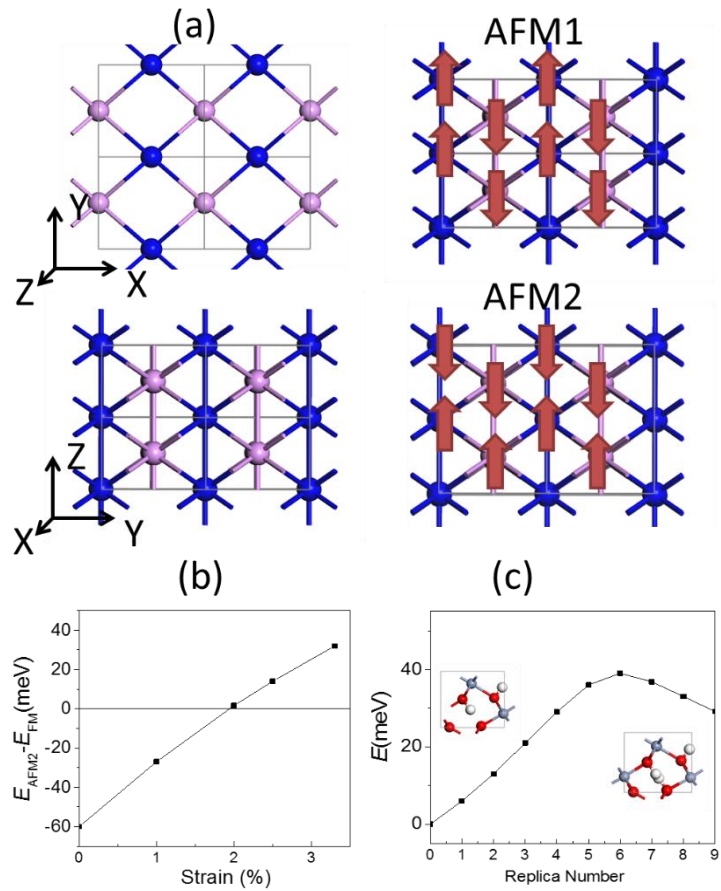

Figure 2 (a) Cr lattice and two AFM spin configuration in CrOOH, where A and B spins are respectively marked by blue and purple spheres, and red arrows denote the spin direction. (b)Dependence of energy difference between AFM2 and FM state on uniaxial strain in z direction. (c)Hopping pathway for one proton to flip to the other side in FE lattice.

Taking β-CrOOH as an example, the FE state is revealed to be lower in energy compared with anti-ferroelectric state (see Fig. S1), and every unitcell contains two Cr atoms and each Cr atom possesses a magnetic moment of $3\mu_B$, which respectively are marked as A (in blue) and B (in purple) in Fig. 2(a). To determine the ground state of spin configuration, we illustrate the Cr spin lattice composed of A and B spins, where every A spin has 8 adjacent B spins and 2 adjacent A spins. The exchange coupling constant between adjacent A-A (or equivalent B-B) and A-B are respectively defined as $J_1$ and $J_2$. For the antiferromagnetic (AFM) configuration where A and B spins are anti-parallel, denoted as AFM1, it will be 8 $J_1$ lower in energy compared with the FM state. If we double the unitcell in z direction, the AFM2 configuration (E-AFM) where adjacent A-A as well as B-B are anti-parallel will be 8 $J_1$ +4 $J_2$ lower in energy compared with the FM state. It turns out that AFM2 is the ground state as we compare the energies of different spin configurations obtained by DFT computations: $E$(AFM1)-$E$(FM)=67.9meV, $E$(FM)-$E$(AFM2)=60.1meV, so $J_1$ and $J_2$ are respectively -8.5 meV and 32.0 meV. We also compared the magnetic frustration states by aligning A spins along z axis and B spins along x or y axis, which are all higher in energy by noncollinear calculations.

The energy required to flip one spin in the AFM2 configuration will be Δ=2$J_2$= 64.0meV. Applying a tensile strain in z direction can greatly change $J_2$ and convert the system from AFM to FM, as shown in Fig. 2(b): upon an uniaxial tensile strain higher than 2%, FM will be the ground state and lower in energy compared with AFM2; as the strain increase to 3.3%, the value $E$(AFM1)-$E$(FM)=31.9meV. A coarse estimation of Curie temperature $T_C$ can be performed simply by using the mean-field theory and Heisenberg model widely used in previous work[40, 41]:

$$T_C = \frac{2\Delta}{3k_B},$$

where Δ=8 $J_1$ +2$J_2$ is the energy required to flip one spin in the FM state with other spins fixed. Upon the strain of 2% and 3.3%, the estimated Curie temperature are respectively 510 and 793K. If we estimate the FE Curie temperature using similar method, as shown in Fig. 2(c), the energy barrier for one proton to flip to the other side in FE lattice (with other protons fixed) is around 39meV by NEB calculations due to the asymmetry of O-H...O: the O...H/O-H distance ratio will be 1.44 Å /1.06 Å and the estimated FE $T_C$ will be 301 K. This barrier will be lower if the nuclear quantum effect is taken into account. However, upon a biaxial strain in XY plane, this barrier will be enhanced and robust at ambient conditions even upon the nuclear quantum effect of protons.

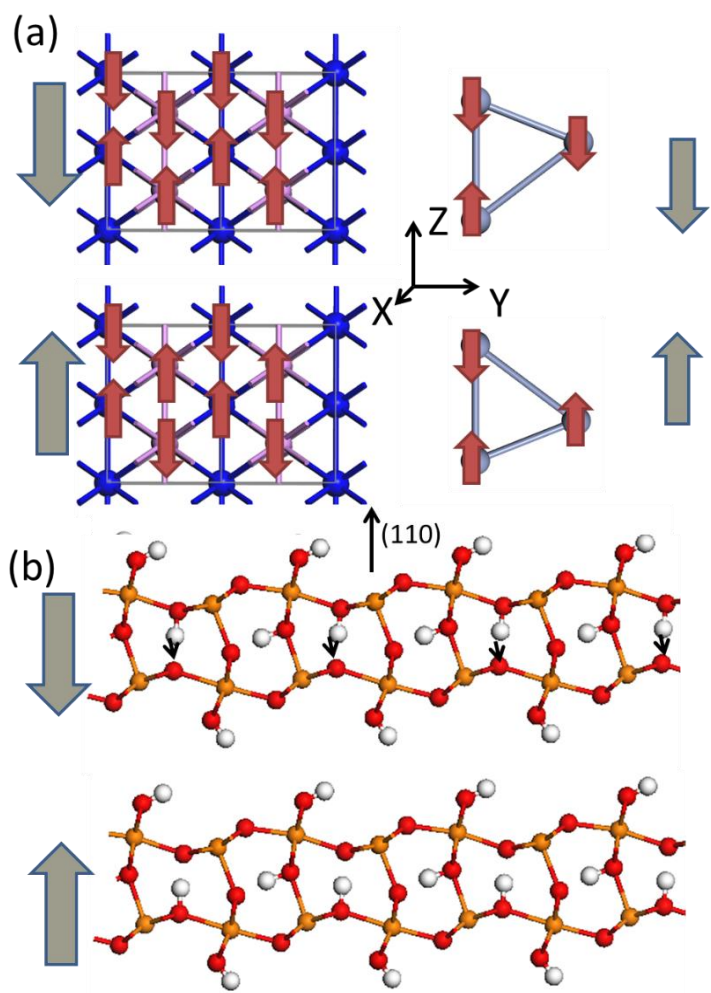

Figure 3 (a) Opposite polarization directions upon AFM2 configurations. (b) FE switching of a thin oxyhydroxide layer isolated from (110) surface. Olive and red arrows denote polarization and spin directions.

It seems that the proton-transfer FE in xy plane is almost completely independent on Cr spins in CrOOH, which should be type-I multiferroics. Meanwhile we note that along the z axis, the AFM2 spin configuration with breaking inversion symmetry may give rise to a weaker switchable polarization: in CrOOH the adjacent Cr-Cr distance will increase by ~0.020 Å when switching from spin parallel state to anti-parallel state due to magnetostriction[42-44], and from I to II in Fig. 3(a), the displacement of Cr ions along the z axis will be around ~0.016 Å. To our computations, this displacement can give rise to a switchable polarization around 1300 $\mu C/m^2$, which is already much higher than the polarizations of prevalent type-II multiferroics (e.g., ~600 $\mu C/m^2$ for $TbMnO_3$)[45]. As a result, CrOOH is type-I multiferroic in xy plane, and type-II multiferroic in z direction where FE and magnetism are coupled. If we suppose an ideal thin layer model of three atomic thickness isolated from (001) surface (see Fig. S3), the ground state will be ferrimagnetic with a net magnetic moment of $3\mu_B$ per unitcell: the Cr spins in the top layer are antiparallel with the spins in the bottom layer, so the spin direction of Cr atoms in the middle layer will determine the direction of both the total magnetization and the polarization. As a result, a vertical polarization of $1.2\times10^{-11}$ C/m is formed. It is known that in traditional ionic FE ultrathin films, FE will disappear below critical film thickness (24 Å in $BaTiO_3$, 12 Å in $PbTiO_3$, for example[37, 38]) due to the depolarizing field. Here the vertical polarization induced by type-II multiferroics is not diminished, and proton-transfer vertical FE of those oxyhydroxides can be even more robust against depolarizing field. For example, as shown in Fig. 3(b), a thin layer of InOOH is isolated from (110) surface so half of the inner O-H...O bonds are almost fixed along the vertical direction, giving rise to a vertical polarization that can hardly be turned to in-plane. The obtained polarization of $3.9\times10^{-11}$ C/m is much higher than previous predicted values in 2D materials (e.g., 0.2 and $1.1\times10^{-11}$ C/m respectively for bilayer BN[29] and functionalized bilayer phosphorene[28]). It is known that the polarization direction of perovskites like $BaTiO_3$ consisting of octahedral $TiO_6$ may be aligned in either of 6 equivalent directions due to the symmetry, so the depolarizing field will symmetrize the ionic bonds or turn the vertical polarization to in-plane in the thin film. For the thin oxyhydroxide layer, however, the direction of O-H...O bonds are firmly aligned when the O atoms at two sides are embedded and almost fixed in the lattice. As a result, a large portion of O-H...O bonds are fixed vertically, which can neither be symmetrized or driven in-plane.

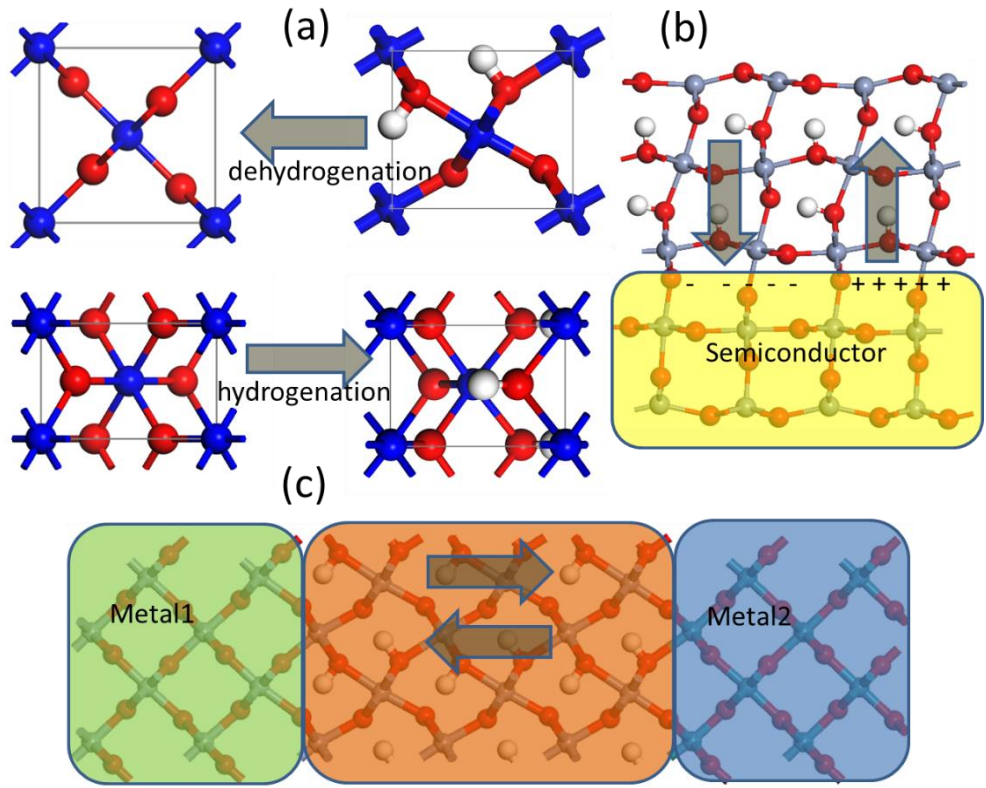

Figure 4 (a) Reversible conversion between metal dioxides and metal oxyhydroxides. (b) Model of FE field effect transistor based on the interface of metal oxyhydroxide and semiconducting metal dioxide. (c) MTJ based on metal oxyhydroxide located between two distinct metal dioxides.

We note the nice match for the lattice constants in Table I, which may be favorable for constructing heterostructure devices. Moreover, those metal oxyhydroxides share similar rutile structures with some metal dioxides (Fig. 4(a)), like $CrO_2$, which is a well-known room-temperature half metal. In a previous experimental report[46] it was shown that supported $CrO_2$ can be reversibly reduced under hydrogen, producing antiferromagnetic CrOOH on titania ($TiO_2$), while reoxidation to $CrO_2$ occurs at temperatures above 520 K in air, under oxygen; and under SCR conditions (NO + $NH_3$ + $O_2$) CrOOH and $CrO_2$ were decomposed at 770 K under argon to antiferromagnetic $Cr_2O_3$. As a result, heterostructure interface composed of different oxyhydroxides and oxides, like CrOOH/$TiO_2$, CrOOH/$CrO_2$, CrOOH/$Cr_2O_3$, can be synthesized via such approach. Herein the CrOOH/$TiO_2$ is a FE/semiconductor interface that can be used for FE field effect transistors[31], as shown in Fig. 4(b). Considering the lattice constants of CrOOH ($|a|$=4.88 Å, $|b|$=4.32 Å) and $TiO_2$ ($|a|$=$|b|$=4.65 Å), the realized CrOOH/$TiO_2$ interface should be the (110) surface, while the (001) surface can match the lattice constants of $Cr_2O_3$ for $CrOOH/Cr_2O_3$ heterostructure. Such nice lattice match will enable perfect epitaxial growth for wider range of applications. If the dioxide substrate is metallic, it may be also favorable to be used as bottom electrode, especially for PFM measurements. FE or multiferroic tunneling junctions can also utilizing the lattice match of CrOOH with $CrO_2$, or other metallic rutile dioxides like $RuO_2$ and $RhO_2$, according to their lattice constants |a| (=|b|) and |c| listed in Table 2. As mentioned above, epitaxial growth of CrOOH on a substrate with relatively slightly larger lattice constants may also strength the ferroelectricity and ferromagnetism of CrOOH. Fig. 4(c) is a design of MTJ composed of $CrO_2/CrOOH/RuO_2$, note that such junction with a high on/off ratio requires two metallic electrodes with significantly different screening length. The difference in screening lengths between $CrO_2$ and $RuO_2$ gives rise to the asymmetry in the electrostatic potential profile that alters an effective barrier height upon FE switching, so two distinct resistances can be obtained. Its on/off ratio is computed by using the nonequilibrium Green's function (NEGF) and Landauer-Buttiker formula[47] implemented in the QuantumWise ATK code[48] (see the model in Fig. S4): the transmission for spin-up and down channel are respectively $1.3 \times 10^{-6}$ and $4.6 \times 10^{-21}$ when the polarization of CrOOH towards the $RuO_2$ side, while changes to $1.2 \times 10^{-7}$ and $5.8 \times 10^{-22}$ when it switches towards the side, so a tunneling electroresistance (TER) as high as 1105% can be achieved. If the electrode $RuO_2$ is substituted by $RhO_2$, the TER will decline to 320%. Meanwhile an intensive magnetoelectric effect can be obtained as the magnetic moment at the interface can vary by 1.05 $\mu_B$ per supercell upon FE switching.

Previously it was predicted that doped-$BaTiO_3/SrRuO_3$ may render electric control of spin injection into FE semiconductors due to the transition between Schottky and Ohmic contacts upon FE switching[49]. Similar mechanism may be utilized in $CrOOH/RuO_2$ or $CrOOH/RhO_2$ interface as long as CrOOH can be doped. Based on the experimental support[46] that $CrO_2$ can be reversibly reduced under hydrogen, the electronic and magnetic structure may be tuned via controlling the density of hydrogen atoms. The hydrogen vacancy in CrOOH can be regarded as *p* doping, which also favors the FM overwhelming AFM. For the structure of $CrOOH_{0.75}$ in Fig. S5(a), FM state will be 33 meV/f.u. lower in energy compared with AFM2. Another report on the synthesis of rutile $Cr_{1-x}Fe_xOOH$ (0<x<1) also reveal the possibility of tuning its properties by doping[50]. For the structure of $Cr_{0.5}Fe_{0.5}OOH$ in Fig. S5(b), our results show that Cr spins (~3 $\mu_B$ per atom) are all antiferromagnetically coupled with Fe spins (~5 $\mu_B$ per atom) in the ground state, so $Cr_{0.5}Fe_{0.5}OOH$ is ferrimagnetic with a magnetic moment of 1.0 $\mu_B$ /f.u..

**Conclusion**

In summary, we provide first-principle evidence of proton-transfer FE in β-GaOOH, InOOH, β-CrOOH and ε-FeOOH. Especially, not only the coexistence of room-temperature magnetism and proton-transfer FE, but also a hitherto unreported combination of type-I and type-II multiferroicity, are predicted in β-CrOOH (guyanaite), rendering a desirable coupling between magnetism and FE. A tiny strain may turn it from antiferromagnetic to ferromagnetic, while the polarization as well as the Curie temperature can also be greatly enhanced. Their atomic-thick thin layer may possess vertical polarization robust against depolarizing field due to the formation of vertical O-H…O bonds that cannot be symmetrized. Due to lattice match for convenient epitaxial growth, heterostructure composed of different rutile oxyhydroxides and metal dioxides, which have been partially synthesized in previous reports, can be constructed for various applications like ferroelectric field-effect transistors and multiferroic tunneling junctions.

**Computational Methods**

Density-functional-theory (DFT) calculations is carried out using the Vienna Ab initio Simulation Package (VASP)[51, 52]. The projector augmented wave (PAW) potentials[53] for the core and the generalized gradient approximation (GGA) in the Perdew-Burke-Ernzerhof (PBE)[54] form for the exchange-correlation functional are applied. The kinetic energy cutoff is set at 530 eV, and the Brillouin zone is sampled by 7×7×9 k points using the Monkhorst-pack scheme. Here following previous models that fit well with experimental data, we checked GGA+U, GGA+D2 and finally adopted GGA+U method on FeOOH where on-site Coulomb and exchange interaction U-J= 5.3 eV is used to treat the d electron states in Fe atoms[11], and pure GGA on CrOOH so the obtained parameters are closer to the values measured in neutron diffraction experiments.[12] Finally the Berry-phase method is used to evaluate crystalline polarization[55]. For the slab calculation, the errors introduced by the periodic boundary conditions can be counterbalanced by setting an electric field to compensate the dipole-dipole interaction between image slabs.

**Conflicts of Interest**

The authors declare no competing financial interest.

**Supporting Information**

The Electronic Supplementary Information is available

**Acknowledgements**

MHW and JML are supported by the National Natural Science Foundation of China (Nos. 21573084, 51721001 and 51431006) and the National Key Research Programme of China (Grant No. 2016YFA0300101). We thank Prof. Ju Li (MIT) for helpful discussions, and Shanghai Supercomputing Center for providing computational resources.